\documentclass[preprint]{aastex}
\input epsf

\newcommand\sss{\scriptscriptstyle\mathsf}

\newcommand\V{{\bf V}}

\newcommand\bnabla{\mbox{\boldmath $\nabla$}}

\begin{document}

\nonumber
\setcounter{equation}{0}

\title{Tuned Finite-Difference Diffusion Operators}
 
\author{Jason Maron\altaffilmark{1}, Mordecai-Mark Mac Low\altaffilmark{2}}
\affil{Department of Astrophysics, American Museum of Natural History,
New York, NY, 10024-5192}

\altaffiltext{1}{jmaron@amnh.org}
\altaffiltext{2}{mordecai@amnh.org}

\begin{abstract}

Finite-difference simulations of fluid dynamics and
magnetohydrodynamics generally require an explicit diffusion operator,
either to maintain stability by attenuating grid-scale structure, or
to implement physical diffusivities such as viscosity or
resistivity. If the goal is stability only, the diffusion must act at
the grid scale, but should affect structure at larger scales as little
as possible. For physical diffusivities the diffusion scale depends on
the problem, and diffusion may act at larger scales as
well. Diffusivity undesirably limits the computational timestep in
both cases.  We construct tuned finite-difference diffusion operators
that minimally limit the timestep while acting as desired near the
diffusion scale.  Such operators reach peak values at the diffusion
scale rather than at the grid scale, but behave as standard operators
at larger scales.  We focus on the specific applications of
hyperdiffusivity for numerical stabilization, and high Schmidt and high
Prandtl number simulations where the diffusion scale greatly exceeds
the grid scale.

\end{abstract}

\section{Introduction}
\label{introduction}

Fluid dynamics simulations usually use explicit diffusion operators,
either to maintain stability or to model physical effects such as
viscosity, resistivity, conductivity, or the diffusion of passive
scalars.  Virtually all astrophysical gas dynamics and MHD simulations
rely on such diffusion operators for stability, physical effects, or
both. We here consider how to design such operators such that they
have the desired behavior at the diffusion scale and larger scales,
while still restricting the numerical timestep as little as
possible. The timestep depends inversely on the strength of
the diffusion
\begin{equation} \label{timestep}
\Delta t = \Delta x^2 / 2 \nu.
\end{equation} Classical diffusion
operators such as Laplacian viscosity ($\nu \nabla^2$), or fourth or
sixth-order hyperdiffusivities ($\nu_4 \nabla^4$ or $\nu_6\nabla^6$), reach
their
maximum values at the grid scale, but act at the larger diffusion
scale where the effective diffusivity is lower. The operators
designed here reach their maximum value at the diffusion scale rather
than at the grid scale, so that they limit the timestep no more than
necessary.

In a spectral code, the diffusive terms are linear and can thus be
handled spectrally without limitation on the timestep. For example,
let a field evolve as $\partial_t \V = A - \nu \bnabla^2 \V,$ where A
denotes the non-diffusive terms.  In Fourier space, $\partial_t
\hat{\V} = A - \nu k^2 \hat{\V}$ The solution, with $A$ constant
throughout the interval $\Delta t$, is
\begin{equation}
\hat{\V}(\Delta t) = \left[ \hat{\V}(0) + \frac{A}{\nu k^2}
(e^{\nu k^2 \Delta t} - 1) \right] e^{-\nu k^2 \Delta t}
\end{equation}
When evolved in Fourier space, the diffusivity operator is stable for
any value of $\nu k^2 \Delta t,$ whereas in physical space, instability
occurs if $\nu k^2 \Delta t > 2$ (restating Eq.~\ref{timestep} in
terms of wavenumber).

However, finite difference codes do have advantages that make them
worth pursuing: they use fewer floating point operations per grid
point; they can be more easily parallelized without the all-to-all
communications required for Fourier transforms; they are not
restricted to periodic boundary conditions; and they handle
discontinuous jumps more robustly.

The Navier-Stokes
equation can include a number of different types of
diffusion operators:
\begin{eqnarray}
\partial_t \V & = & \V \cdot \bnabla \V - \rho^{-1} \bnabla P
+ \nu_2 \bnabla^2 \V - \nu_4 \bnabla^4 \V + \nu_6 \bnabla^6 \V
\nonumber \\
& - & \nu_4^\prime (\partial_x^4 + \partial_y^4 + \partial_z^4) \V
+ \nu_6^\prime (\partial_x^6 + \partial_y^6 + \partial_z^6) \V
- \nu_D D[\V]
 \label{N-S}
\end{eqnarray}
where the $\nu_2$ term is the usual Laplacian physical viscosity, the
$\nu_n$ and $\nu_n^\prime$ terms are $n$th-order hyperviscosities, the
term $\nu_D D(\V)$ is a customized 
diffusion operator.

Either the sixth-order hyperdiffusivity term or the physical diffusivity
can maintain numerical stability. The hyperdiffusivity has been
advocated \citep{bra03} because it preferentially diminishes the
high-wavenumber structure without modifying low-wavenumber structure.
If the problem does require true physical diffusivities, we still want
to consider use of a customized operator.  This would reduce excess
diffusion at scales well below the diffusion scale. Such excess
diffusion limits the timestep without further modifying the solution
as no structure exists at those scales.

To date the focus in the study of extensions to numerical diffusion
has been on such hyperdiffusivities \citep[e.g.][]{borue95,bra03}, as
exemplified by the hyperviscosities described in
equation~(\ref{N-S}). However, the degrees of freedom available in the
finite difference coefficients can be used instead for different goals.

In this paper we describe methods for customizing diffusion operators
that can be used to design operators that protect the timestep while
either minimizing diffusion or reproducing the physical diffusion
operator at low wavenumber as well as possible.  These methods can
also be used to design different diffusion operators for other
purposes.  These methods rely on the tuning techniques used by
\citet{mom08} for improving the high-wavenumber accuracy of
finite-difference derivatives.

In \S~\ref{finite-difference} we summarize the constraints that lead
to the need for tuning finite difference operators.  We then describe
operators suitable for implementing both numerical (\S~\ref{hyper})
and physical (\S~\ref{laplacian}) diffusivities, and we summarize our
results in \S~\ref{summary}.

\section{Tuning Finite Difference Operators}
\label{finite-difference}

Let us consider the question of how to tune a general, symmetric,
finite-difference operator, since all diffusion operators must be
symmetric.  We follow the treatment of the tuning of anti-symmetric
operators such as first derivatives given in \citet{mom08}.  Such
tuning allows us to customize the wavenumber spectrum of the operator
to meet the needs of the problem at hand, rather than relying on simple
analytic forms.  This allows us, for example, in a problem with large
Laplacian diffusivity, to trade small deviations from Laplacian
behavior at low wavenumber for large gains in the timestep, by limiting
the diffusion at high wavenumber.  The deviations from Laplacian at
low wavenumber can be maintained at levels small enough to not affect
realistic simulations. We use both analytic solutions and numerical
optimization to improve the spectral performance of the operators.

As an example of finite difference representations of symmetric
operators we examine second and fourth derivatives.  Define a function
$f_j(x_j)$ on a set of grid points $x_j = j,$ with j an integer. Then
construct a finite difference operator for the second derivative
$f^{[2]}$ by sampling a stencil of grid points with radius $S.$
Without loss of generality, we center the operator on $j=0$ and use a
grid interval of $\Delta x = 1.$ The familiar result for a second
derivative on a radius-1 stencil is
\begin{equation}
\partial_x^2 f(x)|_{x=0} \sim -2 f_0 + f_{1} - f_{-1},
\label{stencil1}
\end{equation}
which is obtained from fitting a
polynomial of degree $2$ to $f_j.$ For a fourth derivative, we can fit
a degree 4 polynomial on a radius-2
stencil, \begin{equation} 
\partial_x^4 f(x)|_{x=0} \sim - 6 f_0 + 4 (f_{-1} + f_{1}) - (f_{-2} + f_{2}).
\label{stencil2}
\end{equation}
In general, a symmetric operator on a stencil of order $S$ can be
represented as
\begin{equation}
m_0 f_0 + \sum_{j=1}^{S} m_j (f_{-j} + f_{j}).
\label{stencil}
\end{equation}

Consider the value of the finite-difference operator at $x=0$ for a
Fourier mode $f = \cos(\pi k x).$ (Sine modes can be ignored because
they don't contribute to the second derivative at $x=0$.)  The
wavenumber $k$ is scaled to grid units so that $k=1$ corresponds to
the maximum (Nyquist) wavenumber $\pi (\Delta x)^{-1}$ expressible on
the grid.  The analytic value for the second derivative is $-\pi^2
k^2,$ whereas the finite difference operator (eq.\ \ref{stencil}) gives
\begin{equation} \label{Dk}
f^{[2]} \sim m_{\sss 0} + 2 \sum_{j=1}^{S} m_j \cos(\pi j k) \equiv -D(k)
\end{equation}
This defines a function $D(k)$ that, when positive, acts as a
diffusion applied to $f(x)$, because the Fourier modes of $f$ evolve
as $\partial_t \hat{f} = - \nu_D D(k) \hat{f},$ where $\nu_D$ is a
viscosity-like parameter that sets the level of diffusion.  
The maximum diffusive timestep is given by the inverse of the
maximum value of $D(k)$ over $0<k<1.$
\begin{equation}
\Delta t < \frac{1}{\nu_D \max[D(k)]}.
\end{equation}
Ideally,
$D(k)$ should scale as $(\pi k)^2$ for $k <
k_d$ and should be constant for $k > k_d.$ 
The focus
of this work is on customizing the form of $D(k)$ so
as to to increase the maximum diffusive timestep, and, in the case of
hyperdiffusion, also minimize low-$k$ diffusion.

\begin{figure}[tbh]
\plotone{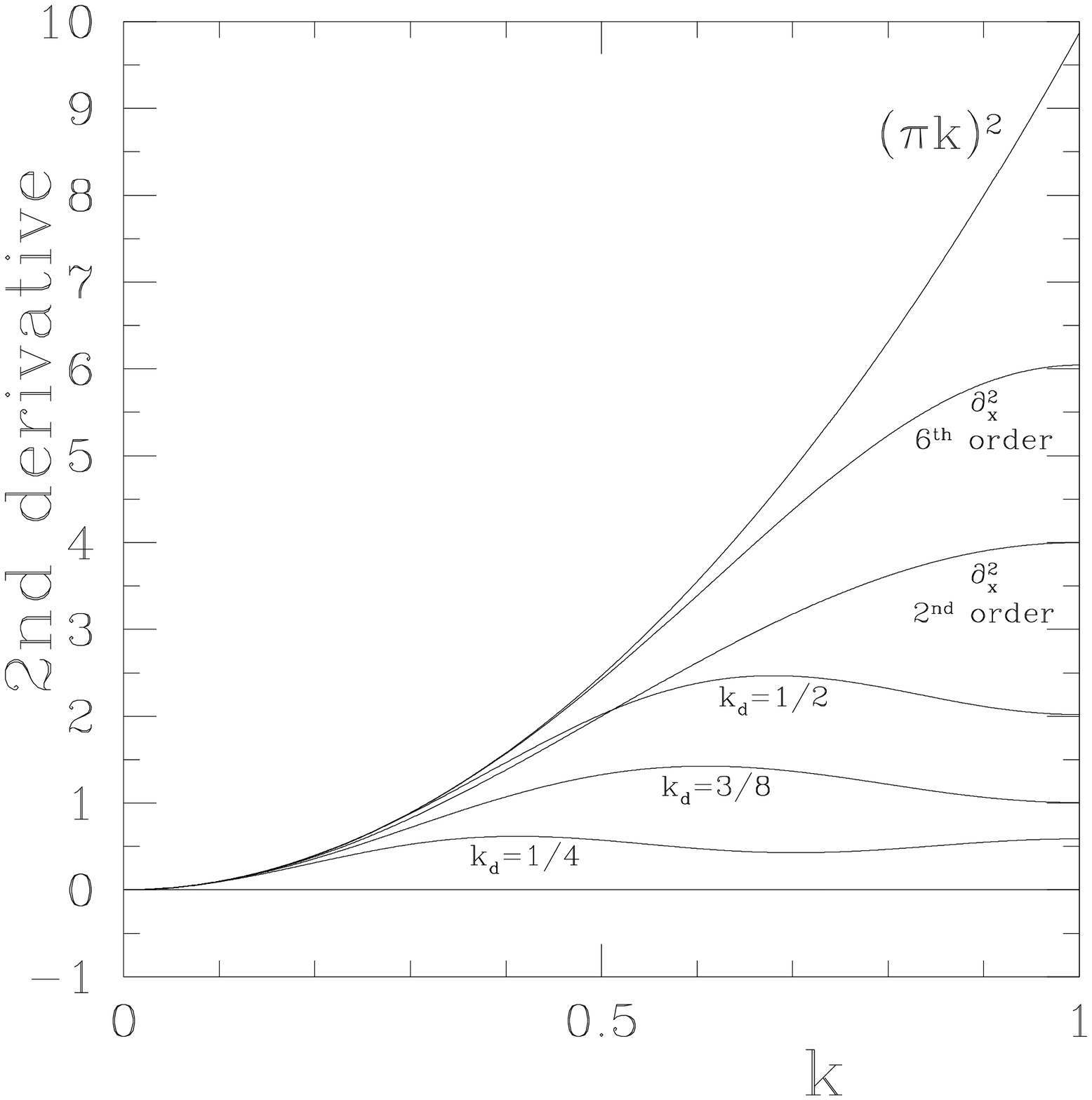} 
\caption{Values of $\partial_x^2 \cos (\pi k x) = (\pi k)^2$, and the
  different finite difference operators having coefficients listed in
  Table~\ref{tablelaplace}.  Two examples of polynomial-based
  difference operators with stencil radii $S = 1$ and highest order two
  and $S = 3$ and highest order six are shown, along with three examples
  of operators tuned with different choices of the beginning of the
  diffusive range $k_d$.}
 \label{figlaplace}
\end{figure}

Figure~\ref{figlaplace} shows $D(k)$ for finite-difference stencils of
radius $S = 1$ (second order) and $S= 3$ (sixth order), in comparison
to the analytic value, demonstrating how higher order more closely
mimics the analytic function.  The coefficients of these functions are
listed in Table~\ref{tablelaplace}. The operator $D(k)$ can be Taylor expanded in the form 
\begin{equation} D(k) = D_0 + D_2
k^2 + D_4 k^4 + D_6 k^6 \ldots.
\end{equation} \label{Dkexp}
An operator that reproduces $\partial^2$ for all $k$ would have $D_2 =
\pi^2$ and all the rest of the coefficients $D_n = 0$ for $n \neq 2$.
The radius-1 stencil (Eq.~\ref{stencil1}) has $D_0 = 0$ and
$D_2=\pi^2$, but the higher order coefficients are unconstrained,
while the radius-2 stencil (Eq.~\ref{stencil1}) sets $D_0=D_4=0$ and
$D_2=\pi^2$.  
\begin{table}
\begin{center}
\caption{Coefficients for timestep-friendly Laplacian diffusion
  operators \label{tablelaplace}} 
\begin{tabular}{lllllll} \tableline \tableline
  &$m_0$ &$m_1$ & $m_2$ &$m_3$ &$m_4$ &$m_5$ \\ 
\tableline
$\partial^2$, $2^{\sss nd}$ order& -2    &1 & \nodata & \nodata & \nodata\\
$\partial^2$, $6^{\sss th}$ order& -2.72222 & 1.5 & -0.15  & 0.01111 &
\nodata  & \nodata \\
$k_d$=1/2 &-1.514721 &0.5692471&0.2524535&-0.0643401& \nodata & \nodata \\
$k_d$=3/8 &-0.9148711&0.2609054&0.2059354&-0.0094052& \nodata & \nodata  \\
$k_d$=1/4 &-0.4334820&0.0752718&0.0696988& 0.0717703& \nodata & \nodata  \\
$k_d$=7/32&-0.2733333&0.0179722&0.0172444& 0.1014500& \nodata & \nodata   \\
$k_d$=3/16&-0.2679784&0.0393498&0.0283651& 0.0304570&0.0358173&  \nodata    \\
$k_d$=9/64&-0.1317918&0.0097316&0.0094934& 0.0088189&0.0081529&0.0296992\\
$k_d$=3/32&-0.0585600&0.0029213&0.0030688& 0.0029842&0.0027554&0.0026748\\
\tableline
\end{tabular}
\tablecomments{The coefficients for a finite difference operator
for $\partial_x^2.$ The 2nd and 6th order entries are for a polynomial
fit on a radius 1 and 3 stencil, respectively.
The ``$k_d$" entries are the tuned
diffusion operators discussed in \S~\ref{laplacian}.These functions are
shown in Figure~\ref{figlaplace}. The last entry, for $k_d=3/32,$
  additionally has $m_6=0.0024189$, $m_7=0.0024876$ and $m_8=0.0099690.$}
\end{center}
\end{table}

\begin{figure}[d4] 
\plotone{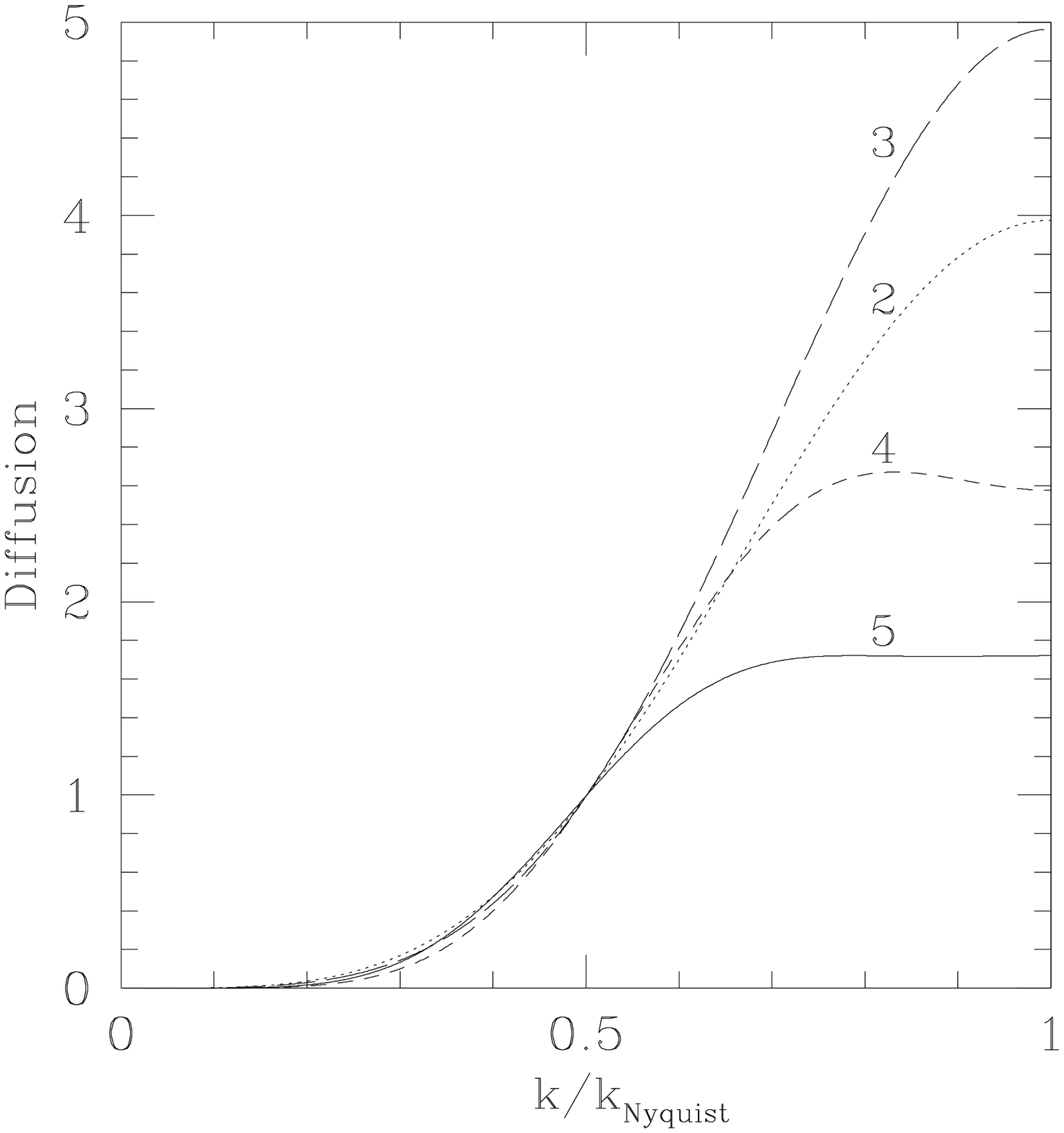}
\caption{Timestep-protecting diffusion functions are compared to
standard hyperdiffusivities. Each of these functions approaches $k=0$
approximately as $k^4,$ and each has been normalized so that $D(1/2) = 1$.
The function labeled ``2'' is the radius-2 stencil finite-difference formula
for $\partial^4$, or in other words, a $\bnabla^4$ hyperdiffusivity.
The function labeled ``3'' is the radius-3 stencil finite-difference formula
for $\partial^4.$ Since it is higher-order, it more faithfully represents
the function $k^4$ for large $k.$ However, this is a liability for the timestep
because it is more vulnerable to a diffusive timestep instability
at $k=1$ than the radius-2 function.
The function labeled ``4'' has a different objective. It is a radius-4 stencil
finite-difference formula for $\partial^4,$ but instead of using the extra
degrees of freedom to represent $k^4$ at higher order, they are used to
minimize $D(k)$ for $k>1/2.$ The function labeled ``5'' has the same
goal as ``4'' implemented with a radius-5 stencil.
\label{figd4}
}
\end{figure}

The usual way to evaluate an operator for a function such as
$\partial^2$ is to fit a maximal order polynomial to the points in the
stencil.  This yields equations that can be inverted to find the $m_j$
coefficients in Equation~\ref{Dk} for stencil radius $S$
\begin{eqnarray} \label{D0}
D_0 & = & - m_0 - 2 \sum_{q=1}^S m_q \\
D_p & = & - \frac{2\pi^p}{p!} (-1)^{p/2} \sum_{q=1}^{S}q^pm_q, \nonumber
\end{eqnarray}
for even $p$. We call such operators polynomial-based
operators. The form of these operators is shown in Figure \ref{figd4}.
We may, however, use the available degrees of freedom in different
ways.  The goal of fitting a high-order polynomial to the derivative
is to have high accuracy at high wavenumber.  However, that may
actually contradict the goal of protecting the timestep while
either minimizing diffusion or reproducing the physical diffusion
operator at low wavenumber. Instead, we can use the available degrees
of freedom to directly address these requirements.

As an example, for a Kolmogorov cascade, the diffusive scale
$\lambda_{\nu}$ and the viscosity $\nu$ scale as $\lambda_{\nu} \sim
\nu^{3/4}.$ The viscosity can be made sufficiently large that
$\lambda_\nu$ is substantially larger than the grid scale, and the
velocity profile will be smooth at smaller scales.  The cascading
energy is elminated at the diffusive scale, so there is no need for
higher diffusivity at smaller scales, yet because of the form of the
Laplacian diffusivity operator, the diffusivity increases all the way
down to the grid scale. This excess diffusivity is unnecessary, and in
fact is a liability because it restricts the timestep.

Laplacian diffusion operators, or steeper operators such as
hyperdiffusivities, rise in amplitude monotonically all the way to the
Nyquist wavenumber $k=1$.  A large value of $D(1)$ requires a small
timestep, but is unnecessary because energy cascading from higher
scales is removed earlier at the diffusion scale $k_d.$ $D(k)$ need
only have enough presence above $k_d$ to diffuse any Fourier modes
that might arise there.  For $k>k_d,$ it can be as large as it is at $k_d$
with no additional timestep penalty.

This allows us to specify a strategically chosen diffusion operator
that satisfies these requirements.  Such an operator should rise
through the diffusion range $k_d$, but then flatten out and merely
remain positive at $k > k_d$. Since this constraint is much less
critical than having controlled diffusion at low $k$, the low-$k$
range of $D(k)$ should receive a higher priority in the optimization
than the high-$k$ range.

The behavior of the diffusion function at $k_d$ critically determines
its effect on the solution. The diffusion must act above both the
Nyquist scale $k_{NY} = 1$, for stability, and at the resolution scale,
to damp modes with wavenumbers too large to be accurately captured by the
finite difference scheme. The resolution scale depends on the
details of the method. However, the common choice of radius-3 stencils
have a resolution scale $k= 1/2$ \citep{mom08},
so we choose in this work to use a
diffusion scale $k_d = 1/2$, and examine diffusion functions
normalized to $D(k_d) = 1$.

We note in passing that if the diffusivity is weak enough to not limit
the timestep, it does not have to be applied
every timestep. The diffusivity can instead be applied once every $N$ timesteps
with a value of of $\nu$ that is $N$ times as large, for reasonable values
of N \citep{mom08}. However, the enhanced diffusivity may then be large enough that
the flat diffusivities described in this paper are required to protect
the timestep. This yields a computational savings from not having
to calculate the diffusion operator every timestep.

In the next two sections we describe how we perform the tuning and
give some useful examples of operators for both hyperdiffusivity and
Laplacian diffusivity.

\section{Timestep-friendly hyperdiffusion}
\label{hyper}

In situations where one wishes to maximize the scale range, and where
the diffusion-scale dynamics don't affect larger scales, one can
fruitfully use a diffusion operator that rises more rapidly with $k$
than a Laplacian.  Writing the diffusive terms from equation \ref{N-S}
in Fourier space,
\begin{equation} \label{eq:diff}
\partial_t \hat{\V} = - \nu_2 k^2 \hat{\V}
- \nu_4 k^4 \hat{\V} - \nu_6 k^6 \hat{\V}
- \nu_4^\prime (k_x^4 + k_y^4 + k_z^4) \hat{\V}
- \nu_6^\prime (k_x^6 + k_y^6 + k_z^6) \hat{\V}
- \nu_d D(k)  \hat{\V}
\end{equation}
one sees that hyperdiffusive terms such as those proportional to
$\nu_4$ and $\nu_6$ have a steeper dependence on the wavenumber $k$
than the Laplacian diffusivity proportional to $\nu_2$. 


However, energy cascading from larger scales dissipates at the
diffusion scale $k_d < 1$, so increasing the diffusion at $k > k_d$
further is unnecessary. A customized diffusion operator $D(k)$ can be
constrained to have a similarly steep $k$ dependence at low $k$, but
to then flatten at the diffusion scale $k_d$, so that the value at the
Nyquist wavenumber $D(1)$ is not markedly higher.  Since the timestep
is limited by the maximum diffusivity on the grid at any scale,
limiting the value of $D$ at small wavenumber protects the timestep.

We note in passing that the $\nu_4$ term in Equation~(\ref{eq:diff})
contains two successive Laplacians and therefore two rounds of finite
differences, whereas terms such as $(\partial_x^4 + \partial_y^4
+ \partial_z^4)$ and $D(k)$ involve only one round of finite
differences, and are therefore favored for their execution speed.
Also, the diffusion function for $\bnabla^4$ has a greater value in
the high-$k$ ``corners'' of Fourier space than $(\partial_x^4
+ \partial_y^4 + \partial_z^4),$ and hence a smaller maximum diffusive
timestep, and so for this reason as well, operators such as
$\bnabla^4$ and $\bnabla^6$ are disfavored.

We begin by considering diffusion operators with a stencil
radius $S=3$, such as are used in the hyperdiffusion implemented in
the Pencil code \citep{bd02}. 
In this case, one can analytically construct a one-parameter family
of functions parameterized by 
the degree of diffusivity $D(1)$ at $k=1$. As before, we take $D(0) =
0$, the diffusion scale $k_d = 1/2$, and normalize $D(k)$ so that
$D(1/2) = 1$. Inverting Equation~\ref{Dk} with these conditions yields
\begin{eqnarray}
m_0 = -\frac{1}{2} - \frac{1}{4} D(1) \\
m_1 = \frac{1}{8} + \frac{7}{32} D(1) \\
m_2 = \frac{1}{4} - \frac{1}{8} D(1) \\
m_3 = -\frac{1}{8} + \frac{1}{32} D(1) \\
\end{eqnarray}
This also implies that 
\begin{equation} \label{d4}
  D_4 = \pi^4\left(\frac18 - \frac{D(1)}{16}\right). \end{equation}  
The magnitude of $D_4$ is inversely
related to the sharpness of the hyperdiffusive filter.  Sharpness of
hyperdiffusivity is usually measured by giving the index of the scaling with wavenumber
$k$ at low $k$. In this example, however, all the functions we present
have $k^4$ scaling at low $k$ and are normalized at $k_d$, so the
lower the value of the coefficient $D_4$ of the $k^4$ term, the
sharper the hyperdiffusivity.

\begin{figure}
\includegraphics[width=6in]{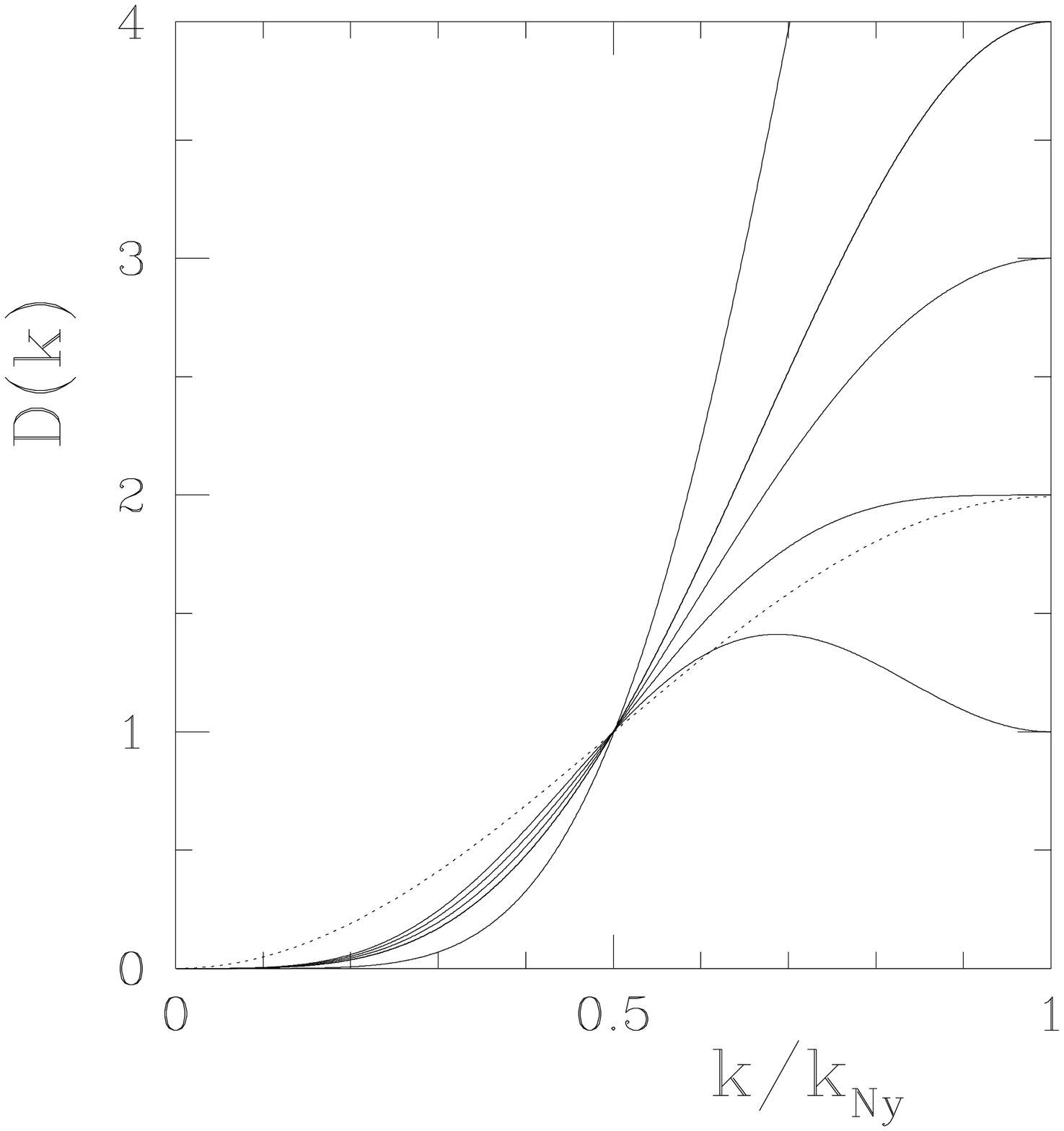}
\caption{The family of diffusion functions $D(k)$ for $1 < D(1) < 8$
for a stencil with radius $S=3$, represented with solid lines. The
dotted line shows the $S=1$ operator for $\partial^2$; the addition of
the two additional parameters in the $S=3$ operator allowed reduction
of diffusivity at both high and low wavenumber while maintaining the
normalization at $k_d$. The $D(1) = 4$
line is identical to the standard $S=2$ hyperdiffusivity. The functions
are all continuous through $D(k_d)$, so the most diffusive above $k_d$
are the least diffusive below it. \label{fig:diff-fcn}
}
\end{figure}

The free parameter $D(1)$ traces the maximum diffusivity, at least in
the regime $D(1)>2,$ as shown in Figure~\ref{fig:diff-fcn}, and thus
determines the timestep.  This can be demonstrated by differentiating
$D(k)$ (Eqs.~\ref{Dk} and~\ref{Dkexp}) and showing that in this
regime, $D'(k) > 0$, so $D(k)$ monotonically increases between $0 < k
< 1$.  For $1 < D(1) < 2,$ the maximum diffusivity is not much greater
than the diffusivity at $k=1.$ For example, for $D(1) = 1.5$, the
maximum diffusivity is $D = 1.63$ at $k = 0.762$.  The useful range
for $D(1)$ is $1 < D(1) < 8$ because the case $D(1)=8$ corresponds to
the operator for $\partial^6,$ which represents the $S=3$
hyperdiffusivity operator that is least diffusive at low
wavenumber. Choosing $D(1) > 8$ results in $D(k) < 0$ for some value
of $k < k_d$.  The choice $D(1) = 8$ corresponds to the standard
hyperdiffusivity used in the Pencil code.  We further find that the
best choice to minimize diffusivity at low $k$ is given by requiring
that $D''(0) = 0.$ If $D''(0)<0,$ then $D(k) < 0$ at low $k$, and
hence unstable, while for $D''(0)>0$, it is more diffusive at low $k$
than it could be.

The goals of minimizing the low-$k$ diffusivity and protecting the
timestep at high $k$ are at odds if one normalizes the diffusion
magnitude at the diffusion scale $D(k_d) = 1$. The tradeoffs can be
seen by considering the behavior of $D(k)$ as $D(1)$ is increased
(Fig.~\ref{fig:diff-fcn}).  Reducing the diffusion for $k < k_d$
requires increasing the diffusion for $k > k_d,$ and vice
versa. Although the values are much larger at high wavenumber, the
percentage changes are actually similar in the two regimes (see
Tab.~\ref{tab:diff-fcn} for the low wavenumber values).

Note that the function with $D(1) = 4$ is the standard hyperdiffusion
with stencil radius $S = 2$, which is the most hyperdiffusive $S=2$ operator.
Adding one extra free parameter by moving to stencil size $S = 3$
allows both low and high wavenumber diffusivity to be tuned, but
we cannot decrease both simultaneously.
However, adding another free parameter by using stencil size $S=4$
does allow both to be decreased 
(Fig.~\ref{figd4}). As an additional example, note that the dotted
line in Fig.~\ref{fig:diff-fcn} shows the $S=1$ diffusivity,
while the $S=3$ result with $D(1)=1.5$ has decreased diffusivity for
both low and high $k$.
Simultaneously decreasing the diffusion for $k<k_d$ and $k>k_d$ while
maintaining a constant diffusion at $k=k_d$ thus clearly requires more
than one free parameter. With two or more free parameters, we can
simultaneously satisfy both goals.


\begin{table}
\begin{center}
\caption{Radius-3 tuned diffusion values \label{tab:diff-fcn}}
\begin{tabular}{lll} \tableline \tableline
$D(1/4)$ & $D(1/3)$ &$D(1)$ \\
\tableline
.124 & .328 & 1.5 \\
.116 & .312 & 2.0 \\
.101 & .281 & 3.0 \\
.086 & .250 & 4.0 \\
.055 & .187 & 6.0 \\
.025 & .124 & 8.0 \\
\tableline
\end{tabular}
\tablecomments{Values of the diffusion function $D(k)$ for
$k=1/4,$ $k=1/3,$ and $k=1,$
for the sequence of radius-3 timestep-friendly hyperdiffusion
functions with varying values of $D(1)$.}
\end{center}
\end{table}

Extending the stencil size to $S > 3$ allows us to further optimize
the diffusion function.
We now have multiple ways in which we could proceed.  We choose to use
numerical optimization to derive 
tuned $S > 3$ operators based on 
the following conditions:
normalize the diffusion spectrum to $D(k_d) = 1$; insist that $0 <
D(k) < \delta$ for some chosen value of the constant $\delta$; set
$D(0) = 0$; and require that $D$ monotonically increase for $k < k_d$.  Within
these constraints, we maximize $D'(k_d)$, which measures the sharpness
of the operator at $k_d$.

To find the operator satisfying these conditions we use a
multiparameter optimization of the coefficients $m_j$ in order to
maximize $D'(k_d)$ within the constraints.  We have developed a novel
Monte Carlo routine to perform the optimization.  It evolves the
solution by testing randomly selected nearby points, selecting the
best among them and iterating with a search radius sensitive to the
speed of improvement of the solution. Because different parameters
have widely varying ranges, we use a logarithmic sampling
distribution.

In Figure~\ref{figd4} we show the resulting optimized diffusivities for
$S=4$ and $S=5$. The coefficients for these operators are given in Table~\ref{tablehyperdiff}. Comparing the $S=2$ hyperdiffusivity to the optimized
$S=4$ operator gives another example of the benefit of taking
advantage of two free parameters.
\begin{table}
\begin{center}
\caption{Coefficients for timestep-friendly hyperdiffusion operators
\label{tablehyperdiff}}
\begin{tabular}{lllllll} \tableline \tableline
  &$m_0$ &$m_1$ & $m_2$ &$m_3$ &$m_4$ &$m_5$ \\ 
\tableline
$\partial^4$, 4th order&1.500000&-1.      & 0.250000& \nodata &
\nodata & \nodata \\
$\partial^4$, 6th order&1.750000&-1.218750& 0.375000&-0.031250&
\nodata & \nodata \\
Tuned, stencil 4       &1.231682&-0.775549& 0.074534&
0.126481&-0.041307& \nodata \\
Tuned, stencil 5       &0.911455&-0.511864&-0.033699& 0.094005& 0.010574&
    -0.014743\\
\tableline
\end{tabular}
\end{center}
\end{table}


\section{Timestep-friendly Laplacian diffusion}
\label{laplacian}

Some applications require such large physical diffusivity that it
becomes the dominant constraint on the timestep.  Examples include
magnetized turbulent flows with separated viscous and resistive scales
so that the magnetic Prandtl number
\begin{equation}
P_m = \frac{\nu}{\eta} 
\end{equation}
is far from unity, where $\eta$ is the resistivity; and turbulence
with a passive scalar such as temperature that diffuses at a scale
different from the viscous scale so that the Schmidt number
\begin{equation}
S = \frac{\nu}{\kappa}
\end{equation}
is far from unity, where $\kappa$ is the
diffusivity of the passive scalar.
In this case, the physical diffusivity operators can also be adjusted
so as to cause less harm to the timestep.

The procedure that we used to generate coefficients for flat Lagrangian
diffusion operators is to specify a value for $k_d,$ and then constrain D so as
to not further increase beyond its value at the diffusion
scale. Specifically, we set
$D(k) < (\pi k_d)^2$ for all $k.$ Within this constraint, we minimize the value
of $(\pi k_d)^2 - D(k)$ over $k > k_d.$ For a radius $S= 3$ stencil, this
procedure works for $k_d$ as low as $7/32.$ Any lower than that and $D(k) <
(\pi k_d)^2$ cannot be satisfied without $D(k)$ taking on a dangerously small
value for some $k > k_d,$ or even worse, becoming negative. However, increasing
the stencil size beyond $S = 3$ allows for flat diffusion operators with
successively lower values of $k_d.$ Such operators are shown in
Table~\ref{tablelaplace}, and Figure~\ref{figlaplace} for $S=3$.

\section{Summary}
\label{summary}

We have presented techniques for customizing diffusion filters with the
goal of either decreasing low-$k$ diffusion, or maximizing the timestep,
or some combination of both. We have given concrete examples
that cover the commonly encountered cases, but since the
requirements for diffusion can be problem dependent, we also
emphasize techniques for customizing general diffusion filters.

Turbulent flows offer a major example of the need for careful choice
of the magnitude of either physical diffusivity or
hyperdiffusivity. The relevant magnitude is that at the diffusion
scale $\nu D(k_d)$, where $k_d$ is chosen to match the spectral
resolution of the numerical scheme \citep{mom08}. There it must be
large enough to absorb the energy from the turbulent cascade reaching
that scale.  The value of $\nu$ is generally set empirically to
satisfy this requirement.

The techniques developed here can also be applied to models with
Prandtl and Schmidt numbers that are large or small compared to unity,
as well as models with diffusive chemistry.

\acknowledgements

We thank J. S. Oishi for useful discussions.
We acknowledge partial support of this work by NSF grant AST06-12724,
and NASA grant NNX07AI74G.

\end{document}